# Performance Analysis of Observation Based Cooperation Enforcement in Ad Hoc Networks

Abeer Ghandar[1], Eman Shabaan[2] and Zaky Fayed[3]

[1] Computer Systems Department, Faculty of Computer and Information Sciences, Ain Shams University
Cairo, Egypt

[2] Computer Systems Department, Faculty of Computer and Information Sciences, Ain Shams University
Cairo, Egypt

[3] Computer Sciences Department, Faculty of Computer and Information Sciences, Ain Shams University
Cairo, Egypt

**Abstract**
Node misbehavior due to selfish or malicious behavior could significantly degrade the performance of MANET because most existing routing protocols in MANET aim to find the most efficient path. Overhearing and reputation based cooperation schemes have been used to detect and isolate the misbehaving nodes as well as to force them to cooperate. Performance analysis has been done for the network traffic using OCEAN over DSR on ns2 while considering the low energy levels for mobile nodes. Throughput, energy level, routing packets and normalized routing overhead are analyzed for OCEAN and normal DSR to show the impact of OCEAN on the overall network performance.
***Keywords:*** *Mobile computing, Protocols, Wireless, DSR, Protocol design and analysis.*

## 1. Introduction

Mobile ad hoc network (MANET) is a group of wireless mobile computers (or nodes), in which nodes cooperate by forwarding packets for each other to allow them to communicate beyond direct wireless transmission range [1]. Ad hoc networks require no centralized administration or fixed network infrastructure such as base stations or access points, and can be quickly and inexpensively set up as needed. They can be used in scenarios in which no infrastructure exists, or in which the existing infrastructure does not meet application requirements for reasons such as security or cost.

These nodes generally have a limited transmission range and, so, each node seeks the assistance of its neighboring nodes in forwarding packets. Specially configured routing protocols are used in order to establish routes between nodes which are further than a single hop. However, cooperation among the nodes is not guaranteed in a real-world network and the presence of misbehaving nodes could degrade the network performance significantly.

MANETs are highly vulnerable to several types of attacks, due to their open medium, lack of centralized monitoring, management point, and lack of strong line of defense. Selfish nodes misbehave to save power or to improve their access to service relative to others [6]. Malicious nodes always attack the network's availability through common techniques such as flooding, black hole and denial of service attacks.

Many contributions to prevent misbehavior have been submitted so far, such as payment schemes for network services, secure routing protocols, intrusion detection, economic incentives and distributed reputation systems to detect and isolate misbehaved nodes [2] [4]. These exiting approaches alleviate some of the problems, but not all.

Despite that these schemes have proved effective. Second-hand reputation systems require nodes in the network to exchange reputation information about other nodes [8] [9]. As a result, exchanging second-hand reputation information opens up a new vulnerability, since nodes may falsely accuse other nodes of misbehaving. If a node observes another node participating incorrectly, it reports this observation to other nodes who then take action to avoid being affected by the misbehavior and perhaps even punish the node by refusing to forward its traffic. Making a decision about whether to believe an accusation requires authenticating and trusting the accusing node. Such trust maintenance could be performed offline or could be bootstrapped during network operations. In the former case, the network requires a priori trust relationships that may not be practical in truly ad hoc networks. In the latter case, bootstrapping trust relationships in ad hoc networks involves significant complexity and risk and may not be reasonable for a very dynamic or short-lived network.

In this paper, OCEAN (Observation-based Cooperation Enforcement in Ad hoc Network) was used in forbidding all kinds of indirect reputation information [1]. A node makes routing decisions based solely on direct observations of its neighboring nodes' exchanges with it. This eliminates most trust management complexity, albeit at a cost of less information with which to make decisions about node behavior [3].

The rest of this paper is organized as follows: Section 2 describes the similar research that has already been done in this area. The detailed protocol is explained in Section 3. The simulation environment is provided in section 4. The results and discussion is in section 5. Finally, section 6 concludes the paper.

## 2. Related Work

Recently, the problem of security and cooperation enforcement has received considerable attention by researchers in the ad hoc network community.

Watchdog and Pathrater was proposed by Marti, Giuli, Lai and Baker [10]. They observed increased throughput in mobile adhoc networks by complementing DSR with a watchdog for detection of denied packet forwarding and a pathrater for trust management and routing policy rating every path used, which enable nodes to avoid malicious nodes in their routes as a reaction. Their approach does not punish malicious nodes that do not cooperate, but rather relieves them of the burden of forwarding for others, whereas their messages are forwarded without complaint. This way, the malicious nodes are rewarded and reinforced in their behavior.

CORE, a collaborative reputation mechanism proposed by Michiardi and Molva [12], also has a watchdog component; however it is complemented by a reputation mechanism that differentiates between subjective reputation (observations), indirect reputation (positive reports by others), and functional reputation (task-specific behavior), which are weighted for a combined reputation value that is used to make decisions about cooperation or gradual isolation of a node. Reputation values are obtained by regarding nodes as requesters and providers, and comparing the expected result to the actually obtained result of a request. Nodes only exchange positive reputation information.

CONFIDANT by S. Buchegger and Jean-Yves Le Boudec [11], also detects misleading nodes by means of observation and more aggressively informs other nodes of this misbehavior through reports sent around the network. Each node in the network hosts a monitor for observations, reputation records for first-hand reports and trusted second-hand reports, trust records to control the trust assigned to the received warnings, and a path manager used by nodes to adapt their behavior according to reputation information.

Researchers have also investigated means of discouraging selfish routing behavior in ad hoc networks through payment schemes [6]. These approaches either require the use of tamper-proof hardware modules or central bankers to do the accounting securely, both of which may not be appropriate in some truly ad hoc network scenarios.

OCEAN is using the same concepts deployed in the Watchdog and Pathrater but it also punishes the selfish and misbehaving nodes in order to force them to cooperate in the network.

## 3. Proposed Scheme

OCEAN is a layer that resides between the network and MAC layers of the protocol stack, and it helps nodes make intelligent routing and forwarding decisions. OCEAN is designed on top of the Dynamic Source Routing Protocol (DSR), although many of its principles may also be useful in other ad hoc routing protocols.

OCEAN divides routing misbehavior into two groups: misleading and selfish. If a node takes part in routes finding but does not forward a packet, it is therefore a misleading node and misleads other nodes. But if a node does not participate in routes finding, it is considered as a selfish node. In order to discover misleading routing behaviors, after a node forwards a packet to its neighbor, it saves the packet in its cache and monitors the neighboring node for a given period of time. It then produces a positive or negative event as its monitoring results in order to update the rating of neighboring node. If the rating is lower than faulty threshold, neighboring node is added to the list of problematic nodes and also added to RREQ as an avoid-list. As a result all traffic will not use this problematic node. This node is given a specific time to return to the network because it is possible that this node is wrongly accused of misbehaving or if it is a misbehaving node, then it must improve in this time period.

OCEAN is composed of five components to discover malicious nodes:
 1. NeighborWatch: observes the behavior of the neighbors of a node.
 2. RouteRanker: holds the nodes ratings for the neighbor nodes.
 3. Rank-based Routing: applies the information from NeighborWatch in the actual selection of routes.
 4. Malicious Traffic Rejection: performs the straightforward rejection of traffic from nodes that are considered misleading.
 5. Second Chance Mechanism: intended to consider the nodes that were previously considered misleading to become useful again.

Ocean attempts to mitigate selfish routing behavior in ad hoc networks. The general idea is to punish nodes for their selfish behavior by rejecting their traffic, in the hopes that this threat will force them to cooperate. OCEAN relies only on direct observations of interactions with neighbors to measure their performance. Every node maintains a chipcount value which acts as a bank balance for every neighbor node. Each node has its own bank to maintain the chipcount values. A node earns chips at every forwarding operation on behalf of the requester node and loses chips with every request. The decision to forward packets for a node is done based on the chipcount for the requester. In order to prevent any deadlocks in the network, the chipcount for all nodes are accumulated with a certain rate.

There are two schemes for incrementing and decrementing chips, optimistic and pessimistic schemes. The optimistic scheme increments the chipcount only when neighbor node accepts the packet. It does not check whether the neighbor node in the route truly forwarded the packets or not. On the other hand, the pessimistic scheme increments the chipcount only when neighbor node is observed to forward the packet.

## 4. Simulation Environment

OCEAN has been deployed on network simulator (ns2) [13] over the DSR protocol. Every point on the following graphs represents an average for 5 simulation runs. Every simulation result is based on the same sent packets for a varying mobile network using the RandomWayPoint mobility model [5]. The energy model of the contributing nodes is considered in the simulation, the initial energy level is low to resemble the nature of sensor networks and different mobile nodes. These simulations model radio propagation using the realistic two-ray ground. The protocol has been analyzed with varying malicious nodes and with varying the faulty threshold. Also, the protocol has been compared to the normal DSR. The conditions for simulation are shown in Table 1. The measurements are done at different pause times to study the different mobility models and the effect of nodes mobility on different parameters of the network taking into consideration the energy of nodes.

Table 1 General Simulation Conditions

| Number of Nodes | 40 |
|---|---|
| Maximum Speed | 20 m/sec |
| Send Rate | 4.0 |
| Packet Size | 512 bytes |
| Simulation Time | 1000 sec |
| Dimensions of Space | 1500 m X 300 m |
| Maximum Connections | 20 |
| Pause Time | 0, 400, 1000 sec |
| Packet Timeout | 1 msec |
| Rating Increment | +1 |
| Rating Decrement | -2 |

Table 2 Values of Ocean for Varying Malicious Nodes

| Faulty Threshold | -40 |
|---|---|
| Second Chance Timeout | 30 |
| Node Rating after second chance | -30 |
| Energy Model | EnergyModel |
| Initial Energy | 5 Joule |
| Transmission Power | 31.32e-3 Joule |
| Receiving Power | 35.28e-3 Joule |
| Idle Power | 712e-6 Joule |
| Sleep Power | 144e-9 Joule |

## 5. Results and Discussion

### 5.1. Varying Malicious Nodes and Pause Time

The first group of results is done with changing the number of malicious nodes using the simulation conditions in Table 2. Generally, it is expected to have the throughput decreasing with the increase in the number of malicious nodes. By analyzing Fig. 1, it is noticed that the static (Pause time = 1000 sec) networks have higher throughput than the highly mobile ones. Also, the results have shown that OCEAN succeeded to maintain the throughput of the network in case of 12% malicious nodes to an average of 81% of the actual throughput and in case of 25% malicious nodes to an average of 68%. This shows the strength of OCEAN.

In Fig. 2, the routing packets are analyzed with different number of malicious nodes. The static networks have the least routing packets because the nodes do not need to discover new routes with every packet sent. On the other hand, the highly mobile networks have very high routing packets because the neighbors of all nodes change very quickly. So, a route discovery request is performed more frequently than the static networks. The graph also shows that the increase in the number of malicious nodes for all mobility models results in decreasing the number of coordinating nodes in the forwarding of packets. This results in using a certain number of nodes in all communications, which results in having almost similar routing packets in a network with 60% or more of malicious nodes.

The final energy level of the network is analyzed as shown in Fig. 3. It is calculated as the sum of the final energy of the nodes divided by the total number of nodes. The graph shows that OCEAN succeeded to consume small amounts of energy to keep the lifetime of the nodes to a maximum level with high throughput at different pause times. Also, the graph shows that the higher the mobility of the network, the more energy is consumed.

5.2 Varying Faulty Threshold

The second group of results is done by changing the faulty threshold, second chance mechanism timeout and node rating after the second chance timeout. This study aims to find the most suitable faulty threshold for every network. The results have been measured with 25% and 50% malicious nodes at different pause times based on the values shown in Table 3. The throughput in this group of results is measured only for any route that is at least 2 hops to see the effect of misleading nodes on the overall throughput.

The small faulty threshold means that the node will be added to the faulty list quickly, even if it is not intentionally malicious. In some cases, link fluctuations or weak signal might cause packet losses. On the other hand, the large faulty threshold means that all nodes will take enough time to use the network resources, whether it is cooperating or misbehaving nodes before being enlisted in the faulty lists of the network nodes.

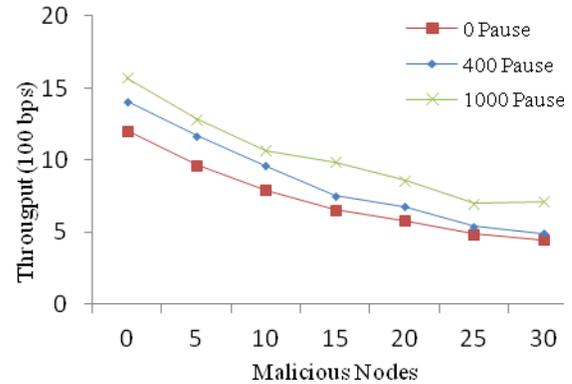

Fig. 1 Throughput of OCEAN with varying malicious nodes and different pause times.

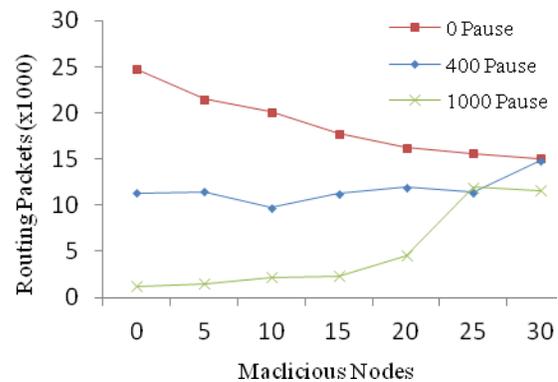

Fig. 2 Routing packets of OCEAN with varying malicious nodes and different pause times.

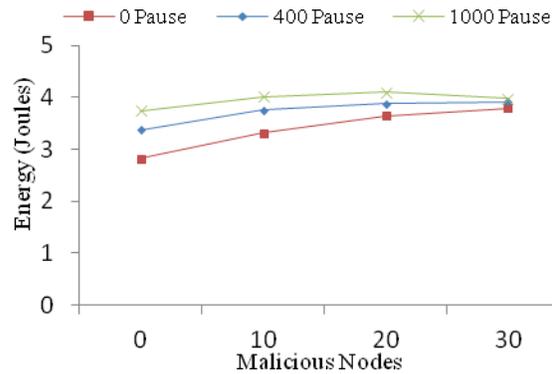

Fig. 3 Energy Level of OCEAN with varying malicious nodes and different pause times.

Since OCEAN depends only on first hand reputation (self reputation) for every node. So, the faulty lists are maintained locally per each node. It proved to have better throughput at low faulty threshold values. Unlike the second hand reputation systems which perform better at higher faulty thresholds. By looking to Fig. 4 and Fig. 5, it is noticed that the throughput was above 80% of original throughput at faulty thresholds less than -50 (low faulty threshold values).

Comparing these results with the second-hand reputation tests done by S. Bansal and M. Baker in [1], it is proved that OCEAN performs better than the second-hand reputation systems in small faulty thresholds. Because the second-hand reputation systems keep a central entity to keep the reputation of all nodes, which needs more time to gather accurate reputation for all nodes of the network. Also, increasing the faulty threshold gives more chance for nodes to be sure that the accused node is truly misbehaving. Hence, it will give more time for malicious nodes to abuse the network resources. In the static networks, the large faulty threshold proved to keep the throughput of the network in the good range. Thus, previous knowledge of the characteristics of the network will help in selecting the most suitable faulty threshold.

By analyzing the routing packets in Fig. 6 and Fig. 7, it is noticed that the routing packets are very high in small faulty thresholds (from 0 to -40). When the faulty threshold is small, the nodes move in and out of the faulty list very quickly. Also, the nodes will do a new route request more frequently because the nodes in the faulty list are updated. These route requests overload the network which consumes the power of the nodes updating the state of the network instead of forwarding packets.

With the thorough analysis of the results, it is clear that choosing a convenient faulty threshold, second chance timeout, new rating for the previously accused nodes is a matter of compromises of the throughput and the actual loading of the cooperating nodes. Also, by further knowledge of the nature of the mobility range of the network, it will help choose the most suitable values for the network parameters.

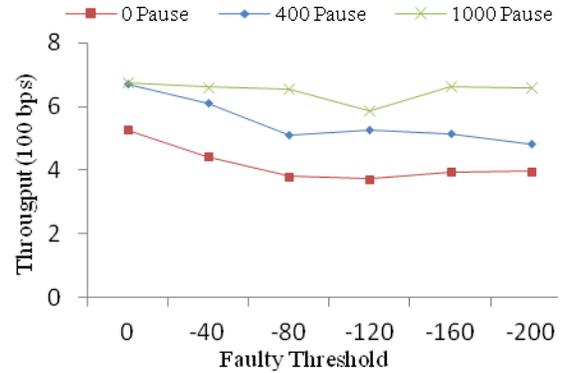

Fig. 4 Throughput at 25% malicious nodes for OCEAN at different faulty thresholds and pause times.

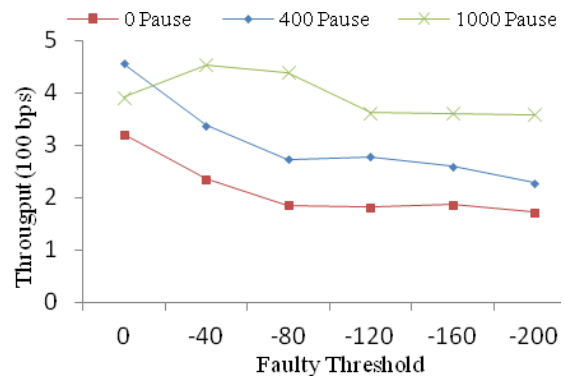

Fig. 5 Throughput at 50% malicious nodes for OCEAN at different faulty thresholds and pause times.

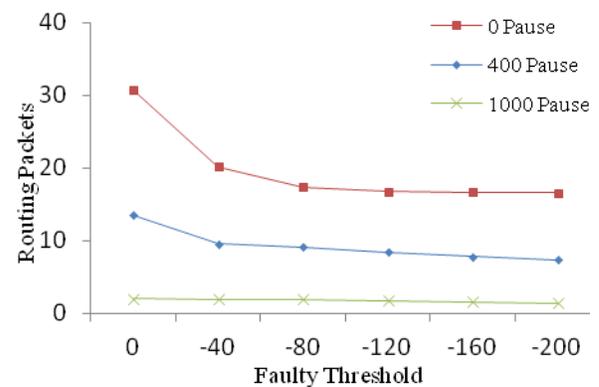

Fig. 6 Routing packets at 25% malicious nodes for OCEAN at different faulty thresholds and pause times.

Table 3 Simulation Conditions for changing faulty threshold and fixed malicious nodes

| Faulty Threshold | Second Chance Timeout Period | Second Chance New Rating |
|---|---|---|
| 0 | 10 | 10 |
| -40 | 30 | -30 |
| -80 | 80 | -70 |
| -120 | 120 | -110 |
| -160 | 160 | -150 |
| -200 | 200 | -190 |

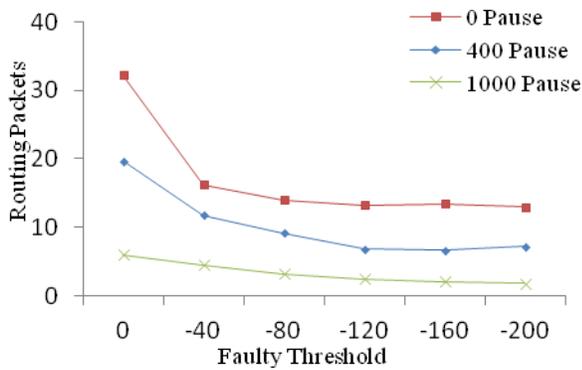

Fig. 7 Routing packets at 50% malicious nodes for OCEAN at different faulty thresholds and pause times.

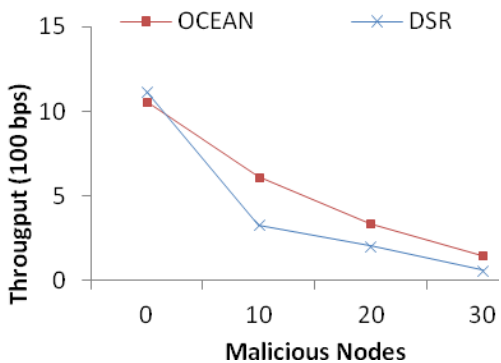

Fig. 8 Throughput for OCEAN and DSR at pause time = 400 ms.

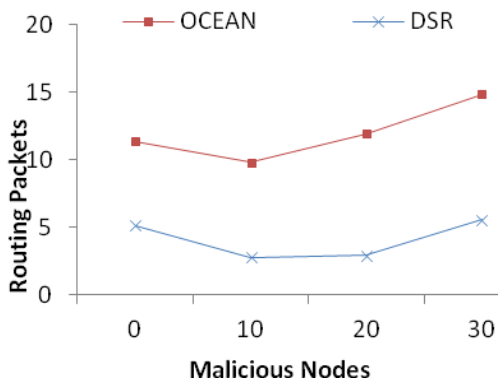

Fig. 9 Routing Packets for OCEAN and DSR at pause time = 400 ms.

### 5.3. Comparing OCEAN to Normal DSR

The third group of results is done by comparing OCEAN to normal DSR without an intrusion detection technique. This group of results aims at finding the effect of adding the OCEAN overhead on the nodes and assessing its impact on the overall network performance. The simulation is done for the OCEAN using the values in Table 1 and Table 2. For the DSR, the malicious nodes are assumed to drop any forwarded packet but they contribute in the route request. The results are compared at medium mobility (Pause Time = 400 ms) with varying number of malicious nodes.

In Fig. 8, the throughput is analyzed in the presence and absence of OCEAN. The graph shows the impact of OCEAN on improving the overall throughput in the presence of up tp 40% of the network misbehaving. Also the graph shows that at zero malicious nodes, the normal DSR has higher throughput than the OCEAN because it does not spend extra time processing nodes ratings.

Fig. 9 shows the routing packets of the network at different number of malicious nodes. OCEAN proved to use more routing packets than normal DSR due to the fact that it does not take the smallest hops path like DSR; on the other hand, it takes the most trusted path which can be longer. Also, it sends packets on the trusted routes only, which excludes other short non-trusted routes. This way, OCEAN uses the good behaving nodes more frequently and ignores the misbehaving nodes to punish them. This overloads the network nodes more.

Fig. 10 represents the delay graph at varying malicious nodes. OCEAN proved to decrease the average delay of packets in the presence of malicious nodes. This is expected because in normal DSR, the dropped packets are re-sent again which increase the overall delay. On the other hand, OCEAN uses the most efficient good behaving route which guarantees that the average delay is at minimum level. It is also noticed that at zero malicious nodes, the delay of normal DSR is lower than that of OCEAN because the extra calculations done by the nodes in OCEAN.

In Fig. 11, the energy level of the network is analyzed for OCEAN and DSR. The network using the OCEAN has less final energy level compared to DSR. Hence, DSR consumes less energy than OCEAN due to the fact that it does not do extra processing on the packets and given that the malicious nodes discarded packets saved the forward energy needed.

## 6. Conclusion

OCEAN succeeded to maintain the throughput of the network to an average of 68% despite having quarter of the nodes of the network misbehaving.

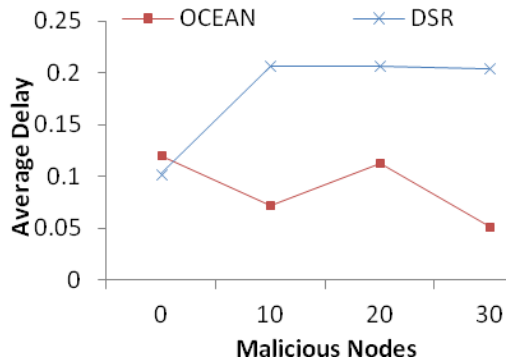

Fig. 10. for OCEAN and DSR at pause time = 400 ms.

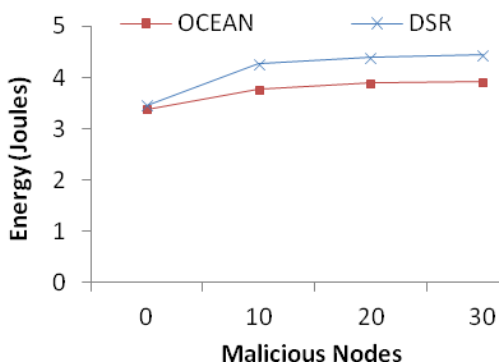

Fig. 11. Energy Level for OCEAN and DSR at pause time = 400 ms.

OCEAN also proved to consume low energy to do the necessary processing. This proves the strength of the protocol in ad hoc networks. The most important strength in this protocol is that it depends on first-hand reputations only. Since it does not need a secondary entity to keep the reputation of each node, which is very adequate with the nature of ad hoc networks. Also, the results discussed in section V show that the parameters of the OCEAN can be configured to an optimum value to achieve the highest throughput and save the network resources with prior knowledge to the properties of the network. The static networks work best with relatively high faulty thresholds while mobile networks work better at lower faulty thresholds.

Also, OCEAN showed that it does not form a huge overhead on the network when compared to normal DSR. On the other hand, it improved the network throughput and average delay and sacrificed the normalized routing load and routing packets to ensure successful delivery of packets and data. OCEAN proved its relevance to the nature of MANETs.

For the future work, OCEAN can be extended on other ad hoc routing protocols and analyzed. Also, further investigations can be done to select the most optimum parameters of OCEAN based on the network characteristics.